\documentclass[useAMS,usenatbib,usegraphicx]{mn2e}

\newcommand{\der}{{\rm d}} 
 \newcommand{\mnras}{MNRAS}
\newcommand{\apj}{ApJ}  
 \newcommand{\apjs}{ApJS} 
\newcommand{\aap}{A\&A}

\newcommand{\R}{{R_{\rm f}}} 
 
\newcommand{\cc}{_{\rm c}}  
\newcommand{\pko}{_{\rm m}}  
\newcommand{\cs}{_{\rm mi}}

 \newcommand{\nest}{^{\rm nn}}

\newcommand{\F}{^{\rm TH}} 
 \newcommand{\h}{_{\rm h}} 
\newcommand{\p}{_{\rm p}} 
\newcommand{\modot}{M$_\odot$\ } 
\newcommand{\ti}{t_{\rm i}} 
\newcommand{\ii}{_{\rm i}}

\newcommand{\beq}{\begin{equation}} \newcommand{\eeq}{\end{equation}}
 \newcommand{\beqa}{\begin{eqnarray}}
\newcommand{\eeqa}{\end{eqnarray}} \newcommand{\lav}{\langle}
\newcommand{\rav}{\rangle} 
\newcommand{\vir}{_{\rm vir}} 
 
 \newcommand{\bb}{}
 
 \newcommand{\ints}{_{\rm int}}

\begin{document}

\title[The Peak Theory] {Fixing a Rigorous Formalism for the Accurate
  Analytic Derivation of Halo Properties}

\author[Juan et al.]{Enric Juan\thanks{E-mail:
    ejrovira@am.ub.es},  Eduard Salvador-Sol\'e, Guillem Dom\`enech and Alberto
  Manrique \\Institut de Ci\`encies del Cosmos.  Universitat de
  Barcelona, UB--IEEC.  Mart{\'\i} i Franqu\`es 1, E-08028 Barcelona,
  Spain}

%% Abstract and keywords

\maketitle
\begin{abstract}
We establish a one-to-one correspondence between virialised haloes and
their seeds, namely peaks with a given density contrast at appropriate
Gaussian-filtering radii, in the initial Gaussian random density
field. This fixes a rigorous formalism for the analytic derivation of
halo properties from the linear power spectrum of density
perturbations in any hierarchical cosmology. The typical spherically
averaged density profile and mass function of haloes so obtained match
those found in numerical simulations.
\end{abstract}



\section{INTRODUCTION}\label{intro}

In the lack of an exact treatment of non-linear structure evolution,
most research in the field of dark matter clustering has been
conducted through $N$-body simulations \citep{FW12}). 

The main difficulty in the analytic derivation of halo properties
comes from the effects of major mergers. For this reason, all efforts
have focused on haloes formed by monolithic collapse or pure
accretion. Nevertheless, as far as virialisation is a real relaxation,
the properties of virialised haloes cannot depend on whether or not
they have suffered major mergers (Salvador-Sol\'e et al.~2012a,
hereafter SVMS).

Following the seminal work by \citet{GG72}, various authors tried to
infer the density profile for haloes emerging by pure accretion from
linear perturbations in the density field at a small cosmic time
$\ti$, assuming spherical collapse and self-similarity (see references
in SVMS). A big step forward was taken when halo seeds were identified
as density maxima (peaks) in the initial Gaussian random field
(\citealt{D70,BBKS}, hereafter BBKS). This led to typical density
profiles in fair agreement with the results of numerical simulations
(\citealt{ARea98,DPea00,As04}). Those solutions were however not yet
fully satisfactory because the typical peak density profile derived by
BBKS is convolved with a Gaussian window and peaks are triaxial and
undergo ellipsoidal collapse. On the other hand, the effects
of shell-crossing during virialisation were not accurately treated.

Other authors concentrated in the halo mass function (MF). Press \&
Schechter (1974) derived it assuming that the seeds of haloes with
mass $M$ at the time $t$ are overdense regions in the Gaussian random
density field at $\ti$ that, smoothed with a top-hat filter at the
scale $M$, have density contrast $\delta$ equal to the critical value
$\delta\cc(t)$ for spherical collapse at $t$. The MF so obtained was
similar to that found in simulations except for a factor
two. \citet{BCEK} corrected this flaw using the excursion set
formalism dealing with the $\delta(M)$ trajectories traced by fixed
points in the initial density field filtered by a sharp k-space window
of varying scale in the presence of an absorbing barrier at
$\delta\cc(t)$ (see also \citealt{ST02} and \citealt{MR10}).
\citet{B88}, \citet{CLM89}, \citet{PH90}, \citet{AJ90}, \citet{bm} and
\citet{ESP} extended this approach to peaks and Manrique \&
Salvador-Sol\'e (1995, hereafter MSS) and \citet{MSS98} developed the
`ConflUent System of Peak trajectories' (CUSP) formalism leading to a
fully consistent analytic derivation of the halo MF from the number
density of non-nested peaks (see also \citealt{H01}).

The CUSP formalism follows from the peak Ansatz inspired by the
spherical collapse that there is a one-to-one correspondence between
virialised haloes with mass $M$ at $t$ and non-nested peaks with
density contrast $\delta(t)$ at the filtering radius
$R(M)$. Unfortunately, these two functions were determined by fitting
the halo MF, which caused the formalism to loose its predicting
power. On the other hand, the validity of the peak Ansatz was not
proved. 

Notwithstanding, this formalism has recently acquired a renewed
interest. As shown by SVMS, it allows one to find the unconvolved
density profile of peaks. Then, taking into account that accreting
haloes develop from the inside out, one can exactly account for the
effects of ellipsoidal collapse and shell-crossing and infer the
typical spherically averaged halo density profile. The used of the
approximated functions $\delta(t)$ and $R(M)$ obtained by MSS could
explain the small departures found in the predicted density profiles
from those found in simulations.

In the present Letter, we justify and accurately fix the halo-peak
correspondence and re-derive the halo density profile and MF. We use
the concordant $\Lambda$CDM cosmology with $\Omega_{\Lambda}=0.73$,
$\Omega_{\rm m}=0.23$, $\Omega_{\rm b}=0.045$, $H_0=0.71$ km s$^{-1}$
Mpc$^{-1}$, $\sigma_8=0.81$ and $n_{\rm s}=1$ together with the BBKS
CDM spectrum with \citet{S95} shape parameter.

\section{The CUSP Formalism}\label{CUSP}

\begin{figure}%[t]
\centerline{\includegraphics[scale=0.46]{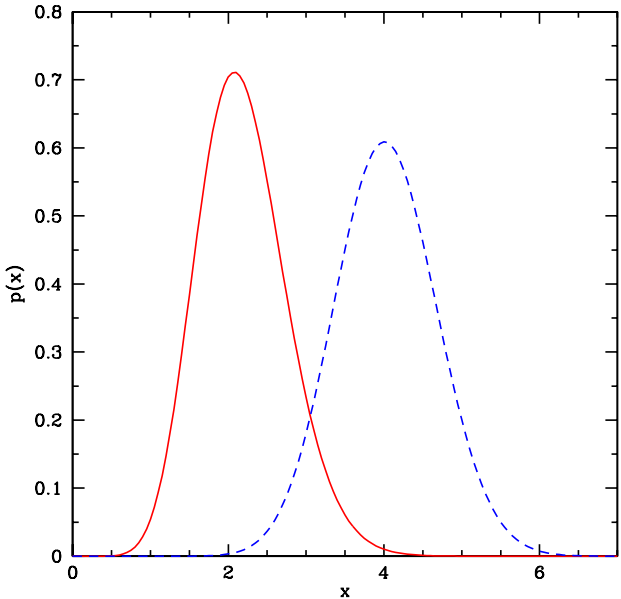}}
\vskip -10pt
\caption{Distribution of curvatures for peaks corresponding to current
  haloes with extreme SO ($\Delta\vir$) masses of $10^{8}$ \modot
  (red solid line) and $10^{16}$ \modot (blue dashed line), for which $\lav (x-\lav
  x\rav)^2\rav^{1/2}/\lav x\rav$ are respectively equal to 0.25 and
  0.16.}
\label{x-distribution}
\end{figure}

Simulations show that virialised haloes form from peaks. Only $15-20$
\% of haloes arise from two nodes \citep{Pea,LP}, which is compatible
with them being currently undergoing a major merger. In fact,
\citet{HP13} found that all {\it virialised} haloes arise from
peaks. However, the one-to-one correspondence between haloes with $M$
at $t$ and peaks at $\ti$ with density contrast dependent only on $t$
as stated in the peak Ansatz seems to be at odds with the idea that
the time of ellipsoidal collapse of peaks depends not only on their
density contrast $\delta$ but also on their ellipticity and density
slope.

But the scatters in ellipticity and in density slope of peaks with
given $\delta$ and $R$ are small compared to the mean values. This can
be seen indeed from the distribution of peak ellipticities (BBKS) and
the result below that the unconvolved density contrast profile
$\delta\p(r)$ of halo seeds is the inverse Laplace transform of the
trajectory $\delta(R)$ they follow when filtered with a Gaussian
filter of varying radius (see eq.~[\ref{dp1}]). The density slope of
seeds, $\der \delta\p/\der r$, is then proportional to the slope $\der
\delta/\der R$ of the peak trajectory,
\beq 
\frac{\partial \delta}{\partial R}= R\,\nabla^2 \delta \equiv -x\,\sigma_2(R)\,R\,,
\label{prime}
\eeq
where $x$ is the peak curvature, that is minus the Laplacian scaled to
the mean value, equal to the second order spectral moment
$\sigma_2$. As the Laplace transform is linear, we then have (see
Fig.~\ref{x-distribution})
\beqa
\frac{\big\lav \big(\frac{\der \rho\p}{\der r}-\big\lav \frac{\der 
    \rho\p}{\der r}\big\rav\big)^2\big\rav^{1/2}}{\big\lav \frac{\der \rho\p}{\der r}\big\rav}
% =\frac{\left\lav \big(\frac{\der \delta}{\der R}-\big\lav \frac{\der
%    \delta}{\der R}\big\rav\big\right)^2\right\rav^{1/2}}{\big\lav \frac{\der
%    \delta}{\der R}\big\rav}\nonumber\\
\approx \frac{\lav (x-\lav x\rav)^2\rav^{1/2}}{\lav x\rav}\ll 1 \,.
\label{ineq}
\eeqa

Thus, if we are interested in the typical properties of haloes, we can
safely assume peaks at $\ti$ with $\delta$ at $R$ having the same
typical ellipticity and density slope. Then, the mass $M$ of
virialised haloes arising at any $t$ from peaks with a fixed value of
$\delta=\delta\cs$ is a function of $R$ alone. And, adopting the halo
mass definition that exactly matches the function $M(R)$, we end up
with the following one-to-one correspondence between haloes with $M$
at $t$ and non-nested peaks at $\ti$ with density contrast $\delta\cs$
at Gaussian-filtering radii $\R$,
\beq 
\delta\cs(t)=\delta\pko(t) \frac{D(\ti)}{D(t)}\,,
\label{deltat}
\eeq
\beq
\R(M,t)=\frac{1}{q(M,t)}\left[\frac{3M}{4\pi\bar\rho(\ti)}\right]^{1/3}\,,
\label{rm}
\eeq
where $\bar\rho\ii$ is the mean cosmic density at $\ti$ and $D(t)$ is
the cosmic growth factor. 

The dependence on $\ti$ on the right of equations
(\ref{deltat})--(\ref{rm}) ensures the arbitrariness of that initial
time. Equation (\ref{deltat}) defines the density contrast
$\delta\pko(t)$ of peaks with $\delta\cs(t)$ at $\ti$ linearly
extrapolated to the time $t$, and equation (\ref{rm}) defines the
radius $q(M,t)$ of halo seeds in units of the radius $\R$ of the
Gaussian filter. As shown by several authors (e.g. \citealt{HP13}),
the density contrast of density perturbations undergoing ellipsoidal
collapse depends on $M$, while in the CUSP formalism it does not. We
note however that in all those works the filter used is top-hat, while
in the CUSP formalism it is Gaussian. This introduces a freedom in
$\R$ associated to a given halo mass $M$ through the function
$q(M,t)$. We can then chose $\delta\cs(t)$ independent of $M$ and let
the radius of the seed in units of $\R$ to depend on $M$.

The use of a Gaussian filter is indeed mandatory for the density
contrast of peaks (with negative values of $\nabla^2 \delta$)
to be always decreasing with increasing filtering radius (see
eq.~[\ref{prime}]), for consistency with the ever increasing mass of
haloes, where $\delta\cs(t)$ is a decreasing function of $t$ and
$\R(M,t)$ an increasing function of $M$. Besides these restrictions,
the functions $\delta\cs(t)$ and $\R(M,t)$ or, equivalently,
$\delta\pko(t)$ and $q(M,t)$ are arbitrary and fix one specific halo
mass definition each. Certainly, the mass definition corresponding to
any given couple of functions $\delta\pko(t)$ and $q(M,t)$ is very
hard to infer and will anyway differ from any usual one, in
general. But, as shown below, we can proceed the other way around:
exactly determine the functions $\delta\pko(t)$ and $q(M,t)$ that
correspond to any desired mass definition.

\section{Fixing the halo-peak correspondence}\label{predictions}

The so-called spherical overdensity (SO) and friends-of-friends (FoF)
mass definitions are the most popular ones. In the former, mostly used
in observational works and numerical studies of the spherically
averaged halo density profile, $\rho\h(r)$, the mass of a halo is that
inside the radius $R\h$ defining an inner mean density
$\bar\rho\h(R\h)$ equal to a fixed overdensity $\Delta$ times the mean
cosmic density,
\beq
\bar\rho\h(R\h)=\Delta \bar\rho(t)\,.
\label{first}
\eeq
$\Delta$ is often taken equal to the cosmology- and time-dependent
virial value $\Delta\vir(t)$ arising from the top-hat spherical
collapse model. This mass definition is from now on referred to as
SO($\Delta\vir$).

But in numerical studies of the MF, the mass of a halo is usually
taken equal to the total mass of its particle members, identified by
means of a FoF percolation finder, with fixed linking
length $b$, in units of the mean interparticle separation. This 
coincides with the mass inside the radius $R\h$ where spheres of
radius $b$ harbour two particles in average \citep{LC94}
\beq
\rho\h(R\h)=\frac{3}{2\pi}\,b^{-3}\bar\rho(t)\,.
\label{second}
\eeq
$b$ is usually taken equal to 0.2 leading to a roughly universal
MF. Such a mass is from now on referred to as FoF(0.2).

In the present Letter, we consider both mass definitions,
SO($\Delta\vir$) and FoF(0.2). This facilitates the comparison with
the results of numerical simulations regarding either the halo density
profile or the MF and illustrates the possibility to apply the same
procedure to any desired mass definition.

\subsection{Spherically Averaged Halo Density Profile}\label{profiles}

\begin{figure}%[t]
%\vskip -10pt
\hskip -4pt
\centerline{\includegraphics[scale=0.46]{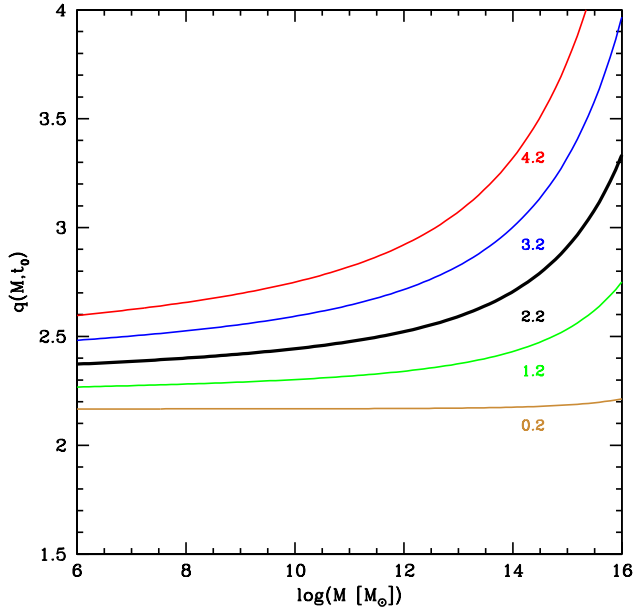}}
\vskip -10pt
\caption{Radius of seeds, in units of Gaussian filtering radius, of
  present haloes with SO($\Delta\vir$) masses (solid lines) and
  FoF(0.19) masses (dashed lines) for the quoted values of
  $\delta\pko(t_0)$. The two kinds of curves fully overlap, but this
  is not the case for any arbitrary mass definition. The thick black
  line is for the value of $\delta\pko(t_0)$ yielding the right
  normalisation of the associated MF.}
\label{qM}
\end{figure}

\begin{figure}
\vskip -10pt
\hskip +7pt
\centerline{\includegraphics[scale=0.46]{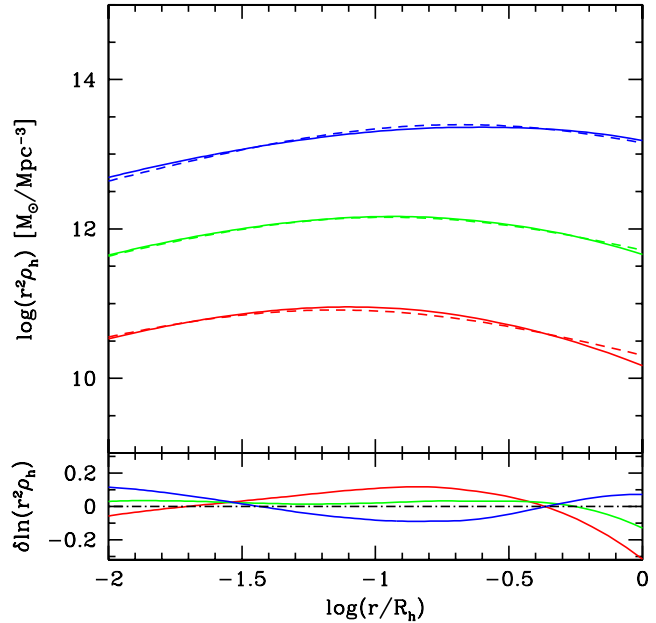}}
\caption{Typical spherically averaged density profiles (solid lines)
  predicted for current haloes with SO($\Delta\vir$) masses
  equal to $5\times 10^{10}$ \modot (red), $5\times 10^{12}$ \modot
  (green) and $5\times 10^{14}$ \modot (blue), compared to the typical
  NFW profiles of simulated haloes (dashed lines) with identical
  masses and the same cosmology according to \citet{ZJMB09}.}
\label{dens}
\end{figure}

As shown in SVMS, during the virialisation of a triaxial halo, shells
cross each other without the crossing of their respective
apocentres. As a consequence, virialised haloes develop from the
inside out, keeping their instantaneous inner structure
unaltered. Then, the radius $r$ encompassing the mass $M$ {\it
  exactly} satisfies the relation\footnote{We are neglecting here for
  simplicity the effects of the cosmological constant (see SVMS for the
  expression accounting for it).}
\beq 
r=-\frac{3GM^2}{10 E\p(M)}\,,
\label{mr2}
\eeq
where $E\p(M)$ is the (non-conserved) total energy of the spherically
averaged seed of the halo progenitor with mass $M$.  Therefore, provided
$E\p(M)$ is known, the relation (\ref{mr2}) can be used to infer the
mass profile $M(r)$ and, by differentiation, the spherically averaged
density profile $\rho\h(r)$ of the final halo.

The energy distribution $E\p(M)$ of the protohalo is given, in the
parametric form, by
\beq E\p(r)=4\pi\!\int_0^{r}\!\!\der \tilde r\, \tilde r^2
\rho\p(\tilde r)
\left\{\!\frac{\left[H_{\rm i}
    \tilde r-v\p(\tilde r)\right]^2}{2}\!-\!\frac{GM(\tilde r)}{\tilde r}
\!\right\}
\label{E1}
\eeq
\beq 
M(r)=4\pi\!\!\int_0^{r}\!\der \tilde r\, \tilde r^2 \rho\p(\tilde r)\,,
\label{M1}
\eeq 
where $\rho\p(r)$ is the (unconvolved) spherically averaged density
profile of the protohalo, $H_{\rm i}$ is the Hubble constant at $\ti$
and
\begin{equation}
  v_p(r)=\frac{2G[M(r)-4\pi r^3\bar\rho\ii/3]}{3H_{\rm_i}r_p^2}
\end{equation}
is, to leading order in the perturbation, the peculiar velocity at $r$
due to the central mass excess.

According to the one-to-one correspondence between haloes and
non-nested peaks, every progenitor of a purely accreting halo arises
from a peak at the corresponding scale $R$. Consequently, the density
contrast $\delta$ at $R$ is but the value at $r=0$ of the spherically
averaged density contrast profile $\delta\p(r)$ of the protohalo
convolved with a Gaussian window of radius $R$,
\beq
\delta(R)=\frac{4\pi}{(2\pi)^{3/2}R^3}
\int_0^{\infty}\der r\, r^2\,\delta\p(r)\,{\rm e}^{-\frac{1}{2}\left(\frac{r}{R}\right)^2}\,.
\label{dp1}
\eeq 

The mean trajectory $\delta(R)$ of peaks tracing the progenitors of a
halo with $M$ at $t$ accreting at the mean rate $\der M/\der t$ and,
hence, resulting with the mean spherically averaged density profile
(remember that accreting haloes grow inside-out) satisfies the
differential equation\footnote{The mean rate $\der M/\der t$
  corresponds to the mean slope $\der R/\der \delta$ rather than to
  mean $\der \delta/\der R$ value. Thus, in equation
  (\ref{dp1}), we should strictly take the inverse of the mean inverse
  curvature, $\lav x^{-1}\rav$, rather than directly $\lav
  x\rav$. But, given the peaked distribution of curvatures, this makes
  no significant difference in the result.}  (see eq.~[\ref{prime}])
\beq 
\frac{\der \delta}{\der R}= -\lav x\rav[R,\delta(R)]\,\sigma_2(R)R\,.
\label{dp2}
\eeq
The mean curvature, $\lav x\rav(R,\delta)$, of peaks with $\delta$ at
$R$ can be calculated for the curvature distribution function given in
MSS, so equation (\ref{dp2}) can be integrated for the boundary
condition $\delta[\R(M,t)]=\delta\cs(t)$ leading to the halo with $M$
at $t$ (eqs.~[\ref{deltat}]--[\ref{rm}]). Once the peak trajectory
$\delta(R)$ is known, equation (\ref{dp1}) becomes a Fredholm integral
equation of first kind for $\delta\p(r)$, which can be solved as
explained in SVMS. Then, bringing the profile
$\rho\p(r)=\bar\rho\ii[1+\delta\p(r)]$ into equations (\ref{E1}) and
(\ref{M1}), we can calculate $E\p(M)$ and, through equation
(\ref{mr2}), obtain the mean spherically averaged density profile
$\rho\h(r)$ for haloes with $M$ at $t$.

The boundary condition $\delta\cs(t)$ at $\R(M,t)$ adopted in SVMS to
solve equation (\ref{dp2}) was derived from equations
(\ref{deltat})--(\ref{rm}) using the {\it approximate} quantities
$\delta\pko(t)$ and $q(M,t)\approx q$ obtained in MSS. This introduced
a small error in the final density profile causing the theoretical
mass at the radius $R\h$, inferred from $M$ {\it according to the
  particular mass definition adopted}, to slightly deviate from this
value $M$. But this suggests the following fully accurate
determination of the function $q(M,t)$ and of the halo density
profile.

Each boundary condition $\delta=\delta\cs(t_0)$ at $R=\R(M_0,t_0)$ for
the integration of equation (\ref{dp2}) gives rise to one peak
trajectory $\delta(R)$ leading to one specific density profile whose
integration out to $r=R_0$ yields a value of the mass different from
$M_0$ in general. Only one particular value of $\R(M_0,t_0)$ or,
equivalently, of $q(M_0,t_0)$ ensures the equality
$M(R_0)=M_0$. Consequently, imposing this constraint, we can find the
desired value of $q(M_0,t_0)$ for any couple of values $M_0$ and
$t_0$. Note that, by changing the value of $\delta\cs(t_0)$ or,
equivalently, of $\delta\pko(t_0)$, the resulting value of
$q(M_0,t_0)$ will change, but neither the solution $\delta(R)$ of
equation (\ref{dp2}) nor the associated final density profile will, so
the particular value of $\delta\pko(t_0)$ used is irrelevant at this
stage. And repeating the same procedure for different masses $M_0$, we
can determine the whole function $q(M,t_0)$ corresponding to any
arbitrary value of $\delta\pko(t_0)$ for any given time $t_0$ (see
Fig.~\ref{qM}).

The mean spherically averaged density profiles so predicted for
current haloes with three SO($\Delta\vir$) masses encompassing the
whole mass range covered in simulations are compared, in Figure
\ref{dens}, to the best NFW fits \citep{NFW97} for simulated haloes
with identical masses obtained by \citet{ZJMB09}. The deviations
observed are typically less than $10$ \%. Only at the outermost radii
in the less massive halo, where the density profile of simulated
haloes is the most uncertain, do they reach 30 \%. Given the absence
of any free parameter in the theory, the agreement found over 4
decades in mass and two decades in radii is remarkable.

The previous result refers to the mean halo density profile. A scatter
is expected arising from that in individual peak trajectories (due to
the scatter in $x$ at each $R$), added to the scatter in the peak
ellipticity and density slope (see Sec.~\ref{CUSP}). In fact, an
``assembly bias'' is foreseen as the peak trajectory $\delta(R)$ of
individual haloes will slightly deviate from the average peak
trajectory and, consequently, the final density profile of individual
haloes and the time at which they reach a given mass fraction will
slightly depend on their mass aggregation history. 

\subsection{Mass Function}\label{MF}

The one-to-one correspondence between haloes and non-nested peaks
implies that the halo MF at $t$, $\partial n(M,t)/\partial M$, coincides, in
comoving units, with the number density of the corresponding
non-nested peaks at $\ti$,
\beq
\frac{\partial n(M,t)}{\partial M}=\,N\nest[\R(M,t),\delta\cs(t)]\,\frac{\partial \R}{\partial M}\,.
\label{mf}
\eeq

Peaks with $\delta\cs$ at scales between $\R$ and $\R + \der\R$ have
density contrasts $\delta$ above $\delta\cs$ at $\R$ and below
$\delta\cs$ at $\R + \der\R$, so they satisfy the condition (see
eq.~[\ref{prime}])
\beq
\delta\cs < \delta \le \delta\cs+x\,\sigma_2(\R)\,\R\,\der\R\,.  
\label{cond}
\eeq
Consequently, the number density of peaks with $\delta\cs$ per
infinitesimal scale around $\R$, $N(\R,\delta\cs)$, is equal to the
number density of peaks per infinitesimal height $\nu\equiv
\delta/\sigma_0$, where $\sigma_0$ is the 0th order spectral moment,
and of curvature $x$, ${\cal N}(\nu,x)$, provided by BBKS, integrated
over all $x$ and over $\nu$ in the range given by the condition
(\ref{cond}).

\begin{figure}%[t]
\vskip -10pt
\hskip 0pt
\centerline{\includegraphics[scale=0.47]{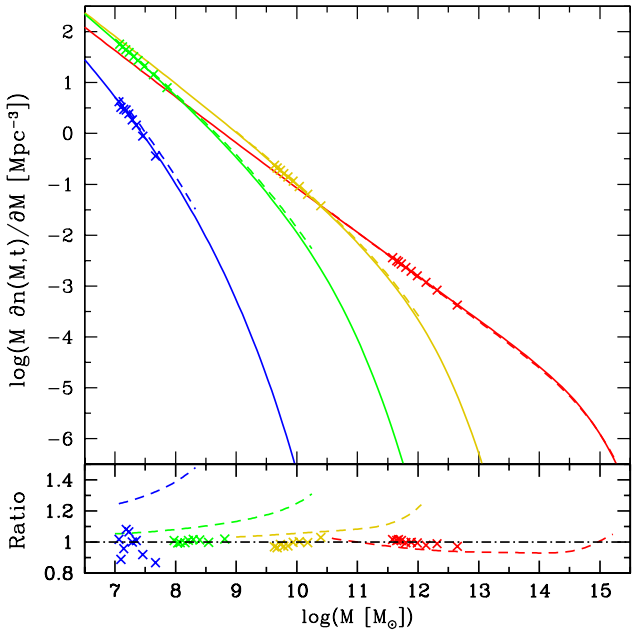}}
\caption{MFs predicted for haloes with FoF(0.2) masses (solid lines),
  compared to \citet{Wea06} analytic fits to the MFs of simulated
  haloes (dashed lines) at $z=20$ (blue lines), 10 (green lines), 5
  (yellow lines) and $0$ (red lines), from left to right. The dashed
  curves cover the ranges analysed in simulations. The ratios in the
  bottom panel are with respect to the theoretical predictions. Points
  are the raw data obtained by \citet{Lea07} in simulations with box
  sizes around $\sim 128(1+z)^{-1}$ Mpc$/h$ giving the best common
  resolution at all $z$'s.}
\label{fm}
\end{figure}

But the density $N(\R,\delta\cs)$ includes all peaks, while
$N\nest(\R,\delta\cs)$ in equation (\ref{mf}) refers only to {\it
  non-nested} ones. Hence, we must correct $N(\R,\delta\cs)$ for
nesting. This is achieved by solving the Volterra integral equation
\beqa
N\nest(\R,\delta\cs)=N(\R,\delta\cs)~~~~~~~~~~~~~~~~~~~~~~~~~~~~~~~~~~~~\nonumber\\
-\frac{1}{\bar\rho\ii}\!\!\int_\R^\infty \!\!\!\der R\bb\, N\nest(R\bb,\delta\cs) M(R\bb,\delta\cs)N\nest(\R,\delta\cs|R\bb,\delta\cs),
\label{nnp}
\eeqa
where the second term on the right gives the number density of peaks
with $\delta\cs$ per infinitesimal scale around $\R$ nested into
non-nested peaks with identical density contrast at any larger
scale. The conditional number density of peaks with $\delta\cs$ per
infinitesimal scale around $\R$ subject to being located in the
collapsing cloud of {\it non-nested} peaks with $\delta\cs$ at $R\bb
>\R$, $N\nest(\R,\delta\cs|R\bb,\delta\cs)$, is the integral over $r$
out to the radius, in units of $q\R$, of collapsing clouds of the
conditional number density of peaks subject to identical conditions
and the additional one of being located at a distance $r$ from a
background peak with $\delta\cs$ at $R$,
$N(\R,\delta\cs|R\bb,\delta\cs,r)$,
\beqa 
N\nest(\R,\delta\cs|R\bb,\delta)=
\frac{3}{C}\int_0^{1}
\der r\, r^2 N(\R,\delta\cs|R\bb,\delta,r)\,.
\label{int}
\eeqa
In equation (\ref{int}), the factor
\beqa
C\equiv\frac{4\pi s^3 N\nest(R\bb,\delta\cs)}{N(\R,\delta\cs)}
%\nonumber~~~~~~~~\\\times
\int_0^{s\ints}\der r\, r^2\,N(\R,\delta\cs|R\bb,\delta\cs,r)\,,
\label{C}
\eeqa
where $s\ints$ is the mean separation, in units of $q\R$, between
non-nested peaks\footnote{This mean separation must be calculated
  iteratively from the mean density (\ref{nnp}). However, two
  iterations starting with $C=1$ are enough to obtain an accurate
  result.}, is to correct for the overcounting of background peaks as
those in $N(\R,\delta\cs|R\bb,\delta\cs,r)$ are not corrected for
nesting. As in the case of the ordinary density of peaks
$N(\R,\delta\cs)$, the conditional density
$N(\R,\delta\cs|R\bb,\delta\cs,r)$ is the integral over all $x$ and
over $\nu$ in the range given by the condition (\ref{cond}) of the
conditional number density of peaks per infinitesimal values of $\nu$
and $x$ subject to being located at a distance $r$ from a background
peak with $\nu'$, ${\cal N}(\nu,x|\nu',r)$, also provided by BBKS.

Using this prescription, every function $q(M,t_0)$ obtained above for each
value of $\delta\pko(t_0)$ will give rise to one possible MF, although
not necessarily satisfying the right normalisation condition
\beq
\bar\rho=\int_0^\infty M(\R)\; N\nest(\R,\delta\cs)\;\der\R\,.
\label{norm}
\eeq
Thus, imposing this constraint, we can determine the right value of
$\delta\pko(t_0)$ and the corresponding function $q(M,t_0)$. And
repeating the same procedure at any time $t$, we can determine the
whole functions $\delta\pko(t)$ and $q(M,t)$.

For FoF(0.2) or more exactly FoF(0.19) masses, the functions
$\delta\pko(t)$ and $q(M,t)$ are found to be identical to those for
SO($\Delta\vir$) masses and take the form
\beq
\delta\pko(t)=\delta\cc(t)\frac{[a(t)]^{1.0628}}{D(t)}\,
\eeq
\beq
q(M,t)\approx \left[Q(M)\frac{\sigma_0\F(M,t)}{\sigma_0(M,t)}\right]^{-2/[n(M)+3]}\,,
\label{qnu}
\eeq 
where $a(t)$ is the cosmic scale factor, $\delta\cc(t)$ is the density
contrast for spherical collapse at $t$, $\sigma_0\F(M,t)$ is the
top-hat 0th order linear spectral moment at $t$ related to $\sigma_0(M,t)$ through
\beq 
\frac{\sigma^{\rm TH}_0(M,t)}{\sigma_0(M,t)}=1- 0.0682\left[\frac{D(t)}{D(t_0)}\right]^{2}\nu\,.
\label{sigmas}
\eeq
$n(M)$ is the effective spectral index at $M$ and $Q(M)$ is defined as
\beq
Q^2(M)=\frac{\int_0^\infty \der x \, x^{2+n(M)}\,W^2_{\rm G}(x)}{\int_0^\infty \der x \, x^{2+n(M)}\,W^2_{\rm TH}(x)}\,,
\label{A}
\eeq 
$W_{\rm TH}(x)$ and $W_{\rm G}(x)$ being the Fourier transforms of the
top-hat and Gaussian windows of radius $x/k$, respectively. Expression
(\ref{qnu}) is approximate as it follows from the more fundamental relation
(\ref{sigmas}), assuming the linear spectrum $P(k)$ equal to a
power-law with spectral index equal to the effective one $n(M)$. This
means that for the CDM spectrum both $n$ and $Q$ depend slightly on
$M$. However, $q(M,t)$ is only needed to calculate $\sigma_0(M,t)$,
which can be readily inferred from the well-known value of
$\sigma_0\F(M,t)$ from the exact relation (\ref{sigmas}).

The MF for FoF(0.19) or SO($\Delta\vir$) masses is compared in Figure
\ref{fm} to the MFs of simulated FoF(0.2) haloes at three redshifts
encompassing the interval studied by \citet{Lea07}. Once again, there
is overall agreement, particularly if we directly compare the
theoretical predictions with the empirical data. Peaks with very low
$\nu$'s will often be disrupted by the velocity shear caused by
massive neighbours. But peaks suffering such strong tides will be
nested, so they will not counted in the MF. This explains why the
theoretical MF is well-behaved even at small masses.

\section{Summary and Conclusions}\label{discuss}

Using simple consistency arguments, we have fixed the one-to-one
correspondence between haloes and non-nested peaks for two popular
halo mass definitions, allowing one to determine the mean spherically
averaged density profile and MF of haloes by means of the CUSP
formalism. The predictions found for SO($\Delta\vir$) and FoF(0.2)
masses in the concordance $\Lambda$CDM model are in good agreement
with the results of numerical simulations.

The CUSP formalism is essentially exact and can be used to derive all
typical halo properties such as the shape and kinematics
\citep{Sea12b}. Moreover it is valid beyond the radius, mass and
redshift ranges covered by simulations and can be applied to cold as
well as warm dark matter cosmologies \citep{Vea12}. It thus has a wide
variety of applications. Furthermore, it allows one to unambiguously
show that accreting haloes grow inside-out and their structure is
independent of their aggregation history \citep{JSM}.

\vspace{0.75cm} \par\noindent
{\bf ACKNOWLEDGEMENTS} \vspace{0.25cm}\par

\noindent This work was supported by the Spanish DGES AYA
2009-12792-C03-01 and AYA2012-39168-C03-02 and the Catalan DIUE
2009SGR00217.  One of us, EJ, was beneficiary of the grant
BES-2010-035483.


\end{document}